\documentclass[aip,graphicx]{revtex4-1}
\usepackage{graphicx}
\usepackage{hyperref}

\draft % marks overfull lines with a black rule on the right

\begin{document}

%\preprint{}

\title{Thermodynamic route of Nb$_3$Sn nucleation: Role of oxygen} %Title of paper



\author{Zeming Sun}
\email[]{zs253@cornell.edu}
%\thanks{}
\affiliation{Cornell Laboratory for Accelerator-Based Sciences and Education, Cornell University, Ithaca, New York 14853, USA}

\author{Darrah K. Dare}
%\email[]{}
%\thanks{}
\affiliation{Cornell Center for Materials Research, Cornell University, Ithaca, New York 14853, USA}

\author{Zhaslan Baraissov}
\affiliation 
{School of Applied \& Engineering Physics, Cornell University, Ithaca, New York 14853, USA}

\author{David A. Muller}
\affiliation 
{School of Applied \& Engineering Physics, Cornell University, Ithaca, New York 14853, USA}

\author{Michael O. Thompson}
\affiliation 
{Materials Science and Engineering, Cornell University, Ithaca, New York 14853, USA}

\author{Matthias U. Liepe}
%\email[]{}
%\thanks{}
\affiliation{Cornell Laboratory for Accelerator-Based Sciences and Education, Cornell University, Ithaca, New York 14853, USA}


\date{\today}

\begin{abstract}
Intermetallic Nb$_3$Sn alloys have long been believed to form through Sn diffusion into Nb. However, our observations of significant oxygen content in Nb$_3$Sn prompted an investigation of alternative formation mechanisms. Through experiments involving different oxide interfaces (clean HF-treated, native oxidized, and anodized), we demonstrate a thermodynamic route that fundamentally challenges the conventional Sn diffusion mechanism for Nb$_3$Sn nucleation. Our results highlight the critical involvement of a SnO$_x$ intermediate phase. This new nucleation mechanism identifies the principles for growth optimization and new synthesis of high-quality Nb$_3$Sn superconductors.

%This finding underscores the significance of incorporating thermodynamic considerations in understanding the early stages of Nb$_3$Sn nucleation. 

\end{abstract}


\maketitle %\maketitle must follow title, authors, abstract and \pacs

\section{Introduction}

Nb$_3$Sn \cite{SunRef35,SunRef36,SunRef37} is a crucial and promising superconductor that finds widespread applications in various modern technologies, including superconducting radio-frequency (SRF) resonators \cite{SunRef4,SunRef5,SunRef1}, high-field magnets \cite{SunRef2}, and emerging electronics such as quantum devices and high-capability detectors \cite{SunRef3}. Of particular interest are SRF resonant cavities that accelerate charged particle beams by coupling RF superconductors with RF waves. These cavities are critical components for modern particle accelerators, with applications spanning from photon science \cite{SunRef6,SunRef7,SunRef8}, high-energy and nuclear physics \cite{SunRef9,SunRef10}, advanced materials discovery \cite{SunRef6,SunRef7}, isotope production \cite{SunRef11}, to quantum computing \cite{SunRef12}.

For SRF applications, Nb$_3$Sn possesses advantages over the conventional material niobium (Nb). Nb$_3$Sn exhibits a high predicted superheating field (400\,mT), twice that of Nb (200\,mT), allowing for large accelerating gradients up to 100\,MV/m \cite{SunRef21}. Nb$_3$Sn has a higher critical temperature ($T_\mathrm{c}$) of 18\,K compared to Nb's 9.2\,K, resulting in lower surface resistance and higher quality factors. The higher $T_\mathrm{c}$ of Nb$_3$Sn enables the replacement of costly helium cryogenics (operated at 2\,K) with commercial cryocoolers (operated at 4.2\,K), thereby facilitating the development of turn-key, compact accelerators \cite{SunRef14}. With excellent capabilities, Nb$_3$Sn-based cavities are poised to advance towards practical accelerator applications \cite{SunRef13}.

\begin{figure} [b!]
\includegraphics[width= \linewidth]{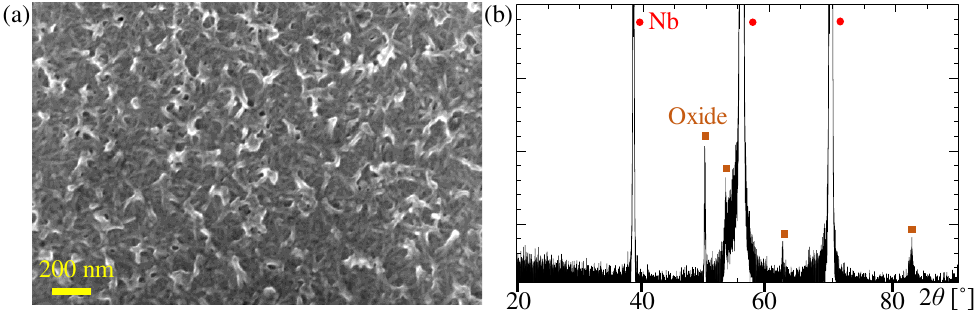}
\caption{\label{SunFig1} (a) SEM surface image and (b) XRD pattern of the anodized Nb.}
\end{figure}

To date, efficient SRF cavities that utilize Nb$_3$Sn films are dominantly produced through tin (Sn) vapor diffusion on Nb surfaces \cite{SunRef13,SunRef22,SunRef23,SunRef24}. Recently, a new process involving Sn electroplating on Nb followed by thermal annealing has been demonstrated \cite{SunRef25}, with other methods, \textit{e.g.}, sputtering, being explored. 

The vapor diffusion process for synthesizing Nb$_3$Sn involves heating Nb in the presence of vaporized Sn at $\sim$\,500\textdegree C for $\sim$\,4\,h for nucleation, followed by heating at $\sim$\,1100\textdegree C for 2\,--\,3\,h for grain growth. To optimize the process, various improvements have been implemented, including switching from native Nb-oxidized surfaces to anodized surfaces, and from the pure Sn source to an additional high-vapor-pressure SnCl$_2$ source. Regardless of these changes, the observation of thick oxides and high subsurface oxygen concentrations in vapor-diffused Nb$_3$Sn \cite{SunRef18,SunRef25} has raised questions about the source of oxygen during nucleation. %These thick oxides are likely to cause losses, while high subsurface oxygen impurities greatly impact performance \cite{SunRef26,SunRef27,SunRef28}. It is therefore crucial to investigate the role of oxygen during nucleation in order to address these practical challenges and improve the process's performance.

\begin{figure} [b]
\includegraphics[width= 90 mm]{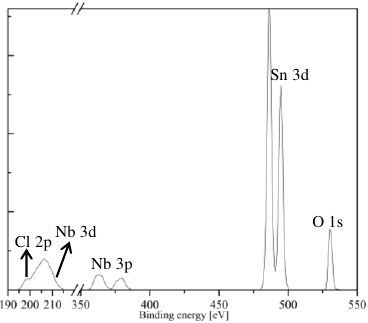}
\caption{\label{SunFig2} XPS spectra for the anodized sample after 300\textdegree C annealing, taken at a depth of 200\,nm. The intensity units are arbitrary.}
\end{figure}

Since the 1980s, the proposed mechanism for intermetallic formation of Nb$_3$Sn has been based on the diffusion of Sn through the grain boundary of Nb or an initially nucleated thin interface of Nb$_x$Sn$_y$. This leads to further growth of Nb$_3$Sn or Nb$_x$Sn$_y$ and slow bulk diffusion to produce Nb$_3$Sn \cite{SunRef29,SunRef30}. Recent atomic simulations have confirmed that grain boundary diffusion in Nb$_3$Sn is significantly faster than bulk diffusion, \textit{e.g.}, 10$^5$ times faster at 1100 \textdegree C \cite{SunRef31,SunRef32}. This diffusion-based mechanism has been widely used to analyze and optimize Nb$_3$Sn growth \cite{SunRef33,SunRef23,SunRef38,SunRef39,SunRef40}.

%Early stage of Nb$_3$Sn nucleation has been explored by using Sn vapor furnace \cite{SunRef38,SunRef39,SunRef40}. 1， this was identified to be Sn which is a pure kinetic problem. This paper identifies the involvement of SnOx thermodynamic routes. 2, the use of vapor technique produces the micrometer size droplet which might be the initial nuclei either Sn or Nb$_3$Sn. However, the identification of these droplets is not verified. EDS on these droplets are not trustable due to oxygen and carbon singles. Cross-sectional TEM is also not trustable while also being challenging. The sample preparation using focused ion beam under 5*10-7 Torr while typically 10-6 Torr the chamber residues give errors on the thin sample. Also, the heavy atoms require relatively large ion polishing energy to produce sufficiently thin sample for TEM imaging. However, this may post-process the surface region. The non-crystalline oxides also provide a challenge to identifying these nuclei. XPS is still the appro technique combined with X-ray diffraction. our sample arch provide a good capping of the interface which is not exposed to air. This increases the chance to probe semi-in situ results. Our analysis is also focused on the interface to showcase the initial nucleation. 

The early stage of Nb$_3$Sn nucleation has been investigated using the Sn vapor process \cite{SunRef38,SunRef39,SunRef40}. This study stands apart from previous research in several ways, challenging existing assumptions. 

Firstly, previous investigations followed the assumption that Sn is the primary component involved in the nucleation process and identified a pure kinetic problem to address non-uniform nucleation issues. However, this work reveals that the thermodynamic route actually involves an intermediate phase of SnO$_x$ in the nucleation process. 

Secondly, vapor-based investigations observed the formation of droplet regions, likely representing the initial nuclei \cite{SunRef38,SunRef39,SunRef40}. However, accurate identification of these droplets remains unverified due to several challenges in characterization. Electron dispersive spectroscopy analysis of these droplets is plagued by unreliable measurements caused by interference from oxygen signals. Cross-sectional transmission electron microscopy imaging poses challenges: sample preparation involving focused ion beam polishing under 5\,$\times$\,10$^{-7}$\,--\,10$^{-6}$\,Torr introduces errors related to the presence of residual oxygen in the chamber; the presence of non-crystalline oxides further complicates the identification of these nuclei. 

Here, we design the sample architectures to ensure a good capping of the interface between the precursor SnCl$_2$ and the substrate, preventing exposure to air. Probing the interface using X-ray photoelectron spectroscopy (XPS) depth profiling under 10$^{-10}$\,--\,10$^{-9}$\,Torr provides semi-\textit{in situ} identification of the nuclei with more reliable results to date.  

%Polishing the sample with heavy atoms necessitates relatively high ion energy, which may inadvertently alter the surface region where nucleation and droplet formation occur.

%This work introduces a fundamentally different mechanism that involves the reaction of SnCl$_2$ to SnO$_x$ and subsequent reduction to metallic Sn during the early nucleation stage (\textit{e.g.}, 300 \textdegree C). This represents a deviation from the previously established diffusion-based mechanism for generating Nb$_3$Sn, and is a critical intermediate step in the process. This creates new opportunities to explore and optimize the material growth process beyond the sole reliance on a diffusion-based mechanism.

\section{Experimental Section}

Three different layouts of SnCl$_2$ / (Nb$_x$O$_y$) / Nb were fabricated: (i) a clean Nb surface treated with hydrofluoric acid (HF) followed by SnCl$_2$ deposition (SnCl$_2$ / Nb); (ii) a native oxidized surface comprising a $\sim$\,7\,nm oxide stack\cite{SunRef18} followed by SnCl$_2$ deposition (SnCl$_2$ / thin, native oxide / Nb); and (iii) an anodized surface with a 60\,nm-thick, porous oxide layer followed by SnCl$_2$ deposition (SnCl$_2$ / thick, porous oxide / Nb). 

High-purity Nb substrates were used with a large grain size of $>$\,50\,mm and a residual-resistivity ratio of $>$\,300. The 1\,$\times$\,1\,cm$^2$ substrates were electropolished using a mixture of 1:9 HF (48\%) and sulfuric (98\%) acids, resulting in an average surface roughness of 40\,nm after a 100\,$\mu$m polishing. The anodization process utilized the electropolished Nb as the anode and Pt as the cathode in a two-electrode control. The solution used was concentrated sodium hydroxide, and the voltage applied was up to 30\,V, producing porous oxides as imaged by scanning electron microscope (SEM) in Fig.~\ref{SunFig1}a.

\begin{table} [t]
  \caption{\label{tbl:1} Binding energies [eV] of relevant motifs in the literature\cite{SunRef15,SunRef16,SunRef17,SunRef18,SunRef19}.}
  \scalebox{1}{
  \begin{tabular}{p{1.5cm}p{4cm}p{4cm}p{3cm}p{3cm}}
    \hline
   Sn 3d$_{5/2}$ & 484\,--\,485~[Metallic] & 486\,--\,487~[SnO/SnO$_2$]& 488~[SnCl$_2$] & 491.2~[Charging] \\
   Nb 3p$_{3/2}$ & 360.5\,--\,361~[Metallic] & 363\,--\,365~[Nb oxides]& 365~[NbCl$_5$] &\\
   O 1s & 529.5\,--\,530.5~[Oxides] & 532\,--\,534~[Hydroxides] & $>$\,537~[Charging] & \\   
   Cl 2p$_{3/2}$ & 198.5\,--199.5~[Chlorides] & & & \\
   \hline
\end{tabular}
}
\end{table}

Afterward, the substrates were transferred to a nitrogen-flown glovebox with O$_2$ and H$_2$O levels below 0.5\,ppm, and all subsequent operations were performed in the glovebox. One type of substrate was soaked in 2\% HF for 30 minutes and dried. SnCl$_2$ dihydrate microparticles were dissolved in methanol at a concentration of 1\,mg/mL. The drop casting technique was employed to deposit 2\,ml of SnCl$_2$ solution onto the substrate surface, followed by solvent evaporation. Uniform coverage of white films was observed on the anodized sample, while it took several attempts to obtain comparable coverage for the other two conditions, which showed poor adhesion.

Following baseline characterizations, the samples underwent annealing in a Lindberg tube furnace under a vacuum of 5\,$\times$\,10$^{-7}$\,torr at 300\textdegree C for 4 hours. While this temperature is lower than the typical nucleation temperature for vapor diffusion (500\textdegree C for 4 hours under $\sim$\,10$^{-6}$\,Torr), it was chosen to capture the early stages of transformation. 

Chemical states, elemental compositions, and phase structures were determined using X-ray photoelectron spectroscopy (XPS) with depth profiling and high-resolution X-ray diffraction (XRD). To accommodate thick films reaching up to 30 $\mu$m, XPS survey scans were performed with sputtering intervals ranging from 1\,nm to 327\,nm, using a Surface Science Instruments SSX-100 ESCA Spectrometer. Monochromatic Al k-alpha X-ray (1486.6\,eV) photoelectrons were collected from an 800\,$\mu$m analysis spot at a 55\textdegree~emission angle under a vacuum of 10$^{-9}$\,Torr. The scan parameters were set to 150\,eV pass energy, 1\,eV step size, and 100\,s/step. For depth profiles, a 4\,kV Ar$^+$ beam with a spot size of $\sim$\,5\,mm was rastered over a 2\,$\times$\,4 mm$^2$ area.

\begin{figure} [b!]
\includegraphics[width= \linewidth]{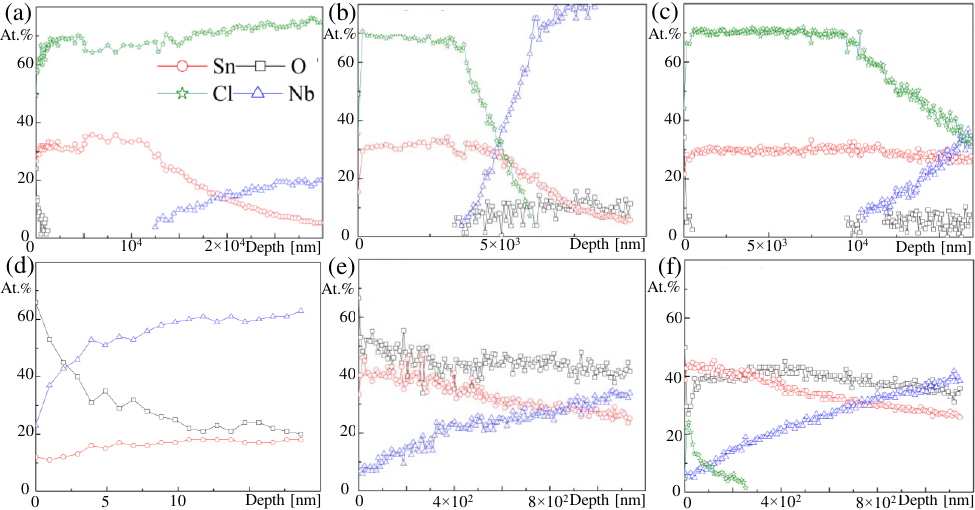}
\caption{\label{SunFig3} Atomic concentrations of Sn, Nb, O, and Cl as a function of depth for the as-deposited (a\,--\,c) and annealed (d\,--\,f) samples. (a,d) Clean Nb, (b,e) native oxidized, and (c,f) anodized. The concentration uncertainty is $<$\,5\%. The depth resolution is affected by the sputtering effect.}
\end{figure}

Phase structures were further analyzed using a Rigaku SmartLab XRD instrument. The X-ray generated by a Cu target was converted into a parallel, monochromatic beam and directed onto the sample surface after passing through a 5\,mm divergence slit. The K$\alpha$ signal with a wavelength of 0.154\,nm was collected by filtering the diffracted beam. The parallel slit analyzer and 5\textdegree~Soller slits were used to resolve high-resolution signals. The 2$\theta$ scan was performed with a step size of 0.0032\textdegree.

\section{Results}

\begin{figure} [b!]
\includegraphics[width= \linewidth]{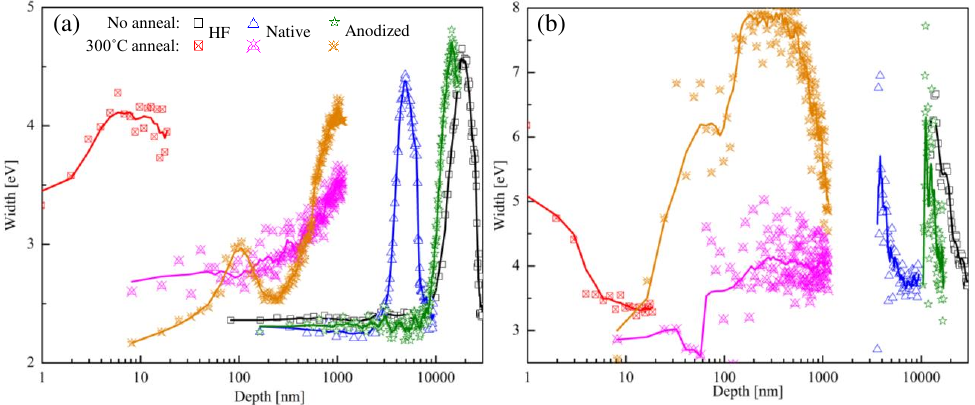}
\caption{\label{SunFig4} XPS peak FWHM of (a) Sn 3d$_{5/2}$ and (b) Nb 3p$_{3/2}$ as a function of depth, with an intrinsic width limit of $\sim$\,2.3\,eV.}
\end{figure}

To gain insight into the structural changes and reactions between multiple elements, XPS spectra of Sn 3d, Nb 3p, O 1s, and Cl 2p (\textit{e.g.}, Fig.~\ref{SunFig2}) were collected as a function of depth for the as-deposited and 300\textdegree C annealed samples. The binding energies are determined with reference to Table~\ref{tbl:1}. To accurately interpret the results, we combine the information from concentration profiles (Fig.~\ref{SunFig3}), FWHM (full-width-at-half-maximum) changes (Fig.~\ref{SunFig4}), and binding energy shifts (Fig.~\ref{SunFig5}), while accounting for the sputtering effects and resolution challenges.

\subsection{As-deposition}
\begin{figure} [b]
\includegraphics[width= \linewidth]{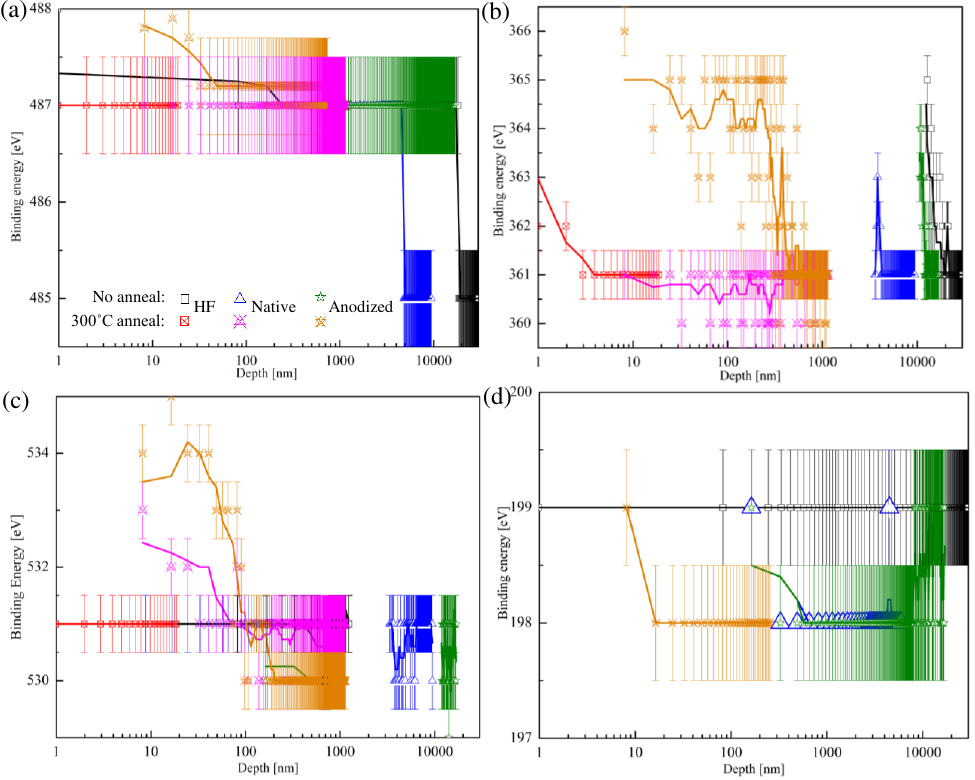}
\caption{\label{SunFig5} XPS binding energies for (a) Sn 3d$_{5/2}$, (b) Nb 3p$_{3/2}$, (c) O 1s, and (d) Cl 2p$_{3/2}$, plotted as a function of depth. The error bars represent a conservative uncertainty estimate of 1\,eV.}
\end{figure}

The oxygen-free interface on clean Nb (Fig.~\ref{SunFig3}a) shows the appearance of NbCl$_x$ and Nb$_x$Sn$_y$ following SnCl$_2$ deposition. Incorporation of these two new Nb and Sn structures results in FWHM increases up to 6.5\,eV and 4.5\,eV, respectively (Fig.~\ref{SunFig4}). The 485\,eV Sn binding energy strongly indicates the generation of Nb$_x$Sn$_y$ alloying upon deposition (Fig.~\ref{SunFig5}a). The 365\,eV Nb binding energy is attributed to NbCl$_x$ (Fig.~\ref{SunFig5}b), as no oxygen signals were detected, ruling out the possibility of Nb oxides that have characteristic peaks in the same positions.

In contrast, the presence of oxygen in the native oxidized and anodized interfaces causes a clear disappearance of chlorine (Fig.~\ref{SunFig3}b,c). The formation of NbCl$_x$ and Nb$_x$Sn$_y$ is still observed, as evidenced by the similar FWHM spikes (Fig.~\ref{SunFig4}) and characteristic binding energies (Fig.~\ref{SunFig5}a,b).

\subsection{After 300\textdegree C annealing}

\begin{figure} [b!]
\includegraphics[width= \linewidth]{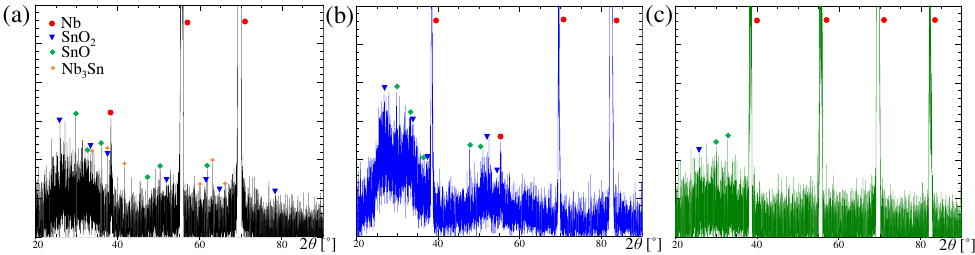}
\caption{\label{SunFig6} High-resolution XRD patterns for the 300\textdegree C annealed samples: (a) clean Nb, (b) native oxidized, and (c) anodized. The data was processed using the GENPLOT smoothing algorithm.}
\end{figure}

Fig.~\ref{SunFig3}d\,--\,f demonstrate the absence of chlorine and the conversion into Sn oxides. To support the phase identification, high-resolution XRD patterns are presented in Fig.~\ref{SunFig6} and referenced in Table~\ref{tbl:2}.   

The clean Nb sample, with a Sn concentration of $\sim$~10\,at.\% and a thinness of $\sim$\,10\,nm, exhibits the lowest amount of SnO$_x$ (Fig.~\ref{SunFig5}a). On the surface, higher-order Nb oxides of $\sim$\,3\,nm are present. Underneath this layer, metallic Nb (Fig.~\ref{SunFig5}b) is mixed with SnO$_x$, resulting in increased Sn FWHM and decreased Nb FWHM (Fig.~\ref{SunFig4}). A low amount of Nb$_x$Sn$_y$ may be present, as suggested in the XRD pattern (Fig.~\ref{SunFig6}a), but the characteristic peak of metallic Sn is not prominent. The rapid decrease of oxygen concentration within $\sim$\,5\,nm for this condition suggests the source of oxygen should be the residual gas in the furnace chamber with a vacuum of 5\,$\times$\,10$^{-7}$\,Torr, which differs from the two oxidized samples where oxygen is evenly distributed at 50\,at.\% throughout the samples.

The native and anodized samples with initial oxides exhibit similarity in the concentration profile (Fig.~\ref{SunFig3}e,f). Both conditions yield significant amounts of SnO$_x$ in addition to Nb$_x$Sn$_y$ nuclei, as evidenced by the persistence of Sn binding energies at 487\,eV associated with SnO$_x$ (Fig.~\ref{SunFig5}a) and the observed diffraction patterns of SnO$_2$ and SnO (Fig.~\ref{SunFig6}b,c). Nb$_x$Sn$_y$ starts to emerge at a depth of $\sim$\,200\,nm, as indicated by the considerably broadened width of both Sn and Nb XPS peaks in Fig.~\ref{SunFig4}. These results provide compelling evidence that SnO$_x$ is a necessary intermediate phase for alloying.

\begin{table} [t]
  \caption{\label{tbl:2}XRD diffraction angles for Nb (Im$\overline{3}$m), SnO$_2$ (P4$_2$/mnm), SnO(P4/nmm), and Nb$_3$Sn (Pm$\overline{3}$n) in the database\cite{SunRef20}.}
  \begin{tabular}{p{2cm}p{2cm}p{2cm}p{2cm}p{2cm}p{2cm}p{2cm}p{2cm}}
    \hline
   Nb & 38.4\textdegree~[110] & 55.4\textdegree~[200]& 69.4\textdegree~[211] & 82.2\textdegree~[220] & & &\\
   SnO$_2$ & 26.4\textdegree~[110] & 33.7\textdegree~[101]& 38.7\textdegree~[111] & 51.5\textdegree~[211] & 54.5\textdegree~[220] & 57.5\textdegree~[002]& 61.5\textdegree~[310]\\
   & 64.3\textdegree~[112] & 71.4\textdegree~[320]& 78.2\textdegree~[321] &&&&\\
   SnO & 29.7\textdegree~[101] & 33.2\textdegree~[110]& 36.8\textdegree~[002] & 47.7\textdegree~[200] & 50.4\textdegree~[112] & 57.2\textdegree~[211]& 61.9\textdegree~[103]\\
   Nb$_3$Sn & 33.7\textdegree~[200] & 37.8\textdegree~[210]& 41.5\textdegree~[211] & 60.2\textdegree~[320] & 62.9\textdegree~[320] & 70.78\textdegree~[400]& \\
     \hline
\end{tabular}
\end{table}

Nevertheless, the anodized sample differs from the native oxidized sample due to the presence of residual SnCl$_2$ on the surface (Fig.~\ref{SunFig3}f and Fig.~\ref{SunFig5}d), resulting in a higher Sn binding energy (Fig.~\ref{SunFig5}a) and a width spike in the Sn peak (Fig.~\ref{SunFig4}a) at the $<$\,200\,nm surface. Within the same depth range, the anodized oxides persist and convert to a larger quantity of higher-order oxides, as evidenced by the 365\,eV binding energies (Fig.~\ref{SunFig5}b) and the largest Nb peak width with all charge states (Fig.~\ref{SunFig4}b). The adoption of anodization for vapor diffusion produces a smoother Nb$_3$Sn surface. Our data suggest that the underlying mechanisms involve two aspects: (i) a larger and more uniform adsorption of SnCl$_2$ due to the porous oxide structures, as shown in Fig.~\ref{SunFig3}c, and (ii) a prolonged reaction time with additional SnCl$_2$ source at a later stage and additional Nb oxides for producing the intermediate SnO$_x$ phases necessary for generating the intermetallic alloy.  

\section{Discussion}

The early stage of Nb$_x$Sn$_y$ nucleation (\textit{e.g.}, at 300 \textdegree C) crucially determines the quality of the final Nb$_3$Sn product, especially in applications involving vapor diffusion in SRF cavities. Achieving a smooth surface in Nb$_3$Sn relies on the uniformity of the initial nuclei, which serve as a seed layer for promoting uniform grain growth. Moreover, Nb$_{3-n}$Sn has stoichiometric boundaries of 18\,at.\% and 25\,at.\% Sn, with the desired stoichiometry of 25\,at.\% Sn corresponding to the critical temperature limit of 18\,K. Correcting the nucleation from an initial 18\,at.\% stoichiometry to the desired 25\,at.\% is challenging. Previous nucleation studies assumed Sn to be the main component involved, considering it to be a purely kinetic problem dependent on the availability of sufficient Sn supply \cite{SunRef38,SunRef39,SunRef40}. However, our observations challenge the application of this assumption to the \textit{nucleation} process, while it remains unclear to what extent pure Sn diffusion, without the assistance of oxygen, contributes to subsequent grain growth.  

Our findings reveal the critical and dominant involvement of SnO$_x$ as a significant intermediate step during the early nucleation stage of Nb$_x$Sn$_y$. Our experimental design has addressed several confusions related to nucleation during vapor diffusion while minimizing uncertainties associated with previous studies. The results directly clarify previous observations of nuclei-like droplet features, providing clear evidence that these features are not Sn droplets, but rather SnO$_x$ entities that contain converted Nb$_x$Sn$_y$. This rectifies several proposed reactions that aimed to balance the formulas with Sn as the end product, \textit{e.g.}, in Reference [\cite{SunRef40}]. In future work, thermodynamic considerations should be prioritized, and models for SnO$_x$/Nb and SnO$_x$/NbO$_x$ should be constructed in experiments and simulations.

These findings also explain the high oxygen content that remains in the final Nb$_3$Sn product. Moreover, while investigating grain growth in Nb$_3$Sn is beyond the scope of this study, the connection between the high oxygen content in Nb$_3$Sn and the involvement of SnO$_x$ implies that oxygen is highly likely involved in the grain growth process. This point should be further verified. 

Furthermore, in addition to providing insights into vapor diffusion, this work has the potential to open up new avenues for generating Nb$_3$Sn, moving beyond vapor diffusion.

\section{Conclusion}

To summarize, this work demonstrates the thermodynamic route for Nb$_3$Sn nucleation. Our results reveal the crucial role of a SnO$_x$ intermediate phase, leading to alloy nucleation. This mechanism is particularly pronounced at the native oxidized and anodized interfaces, as well as clean Nb interfaces due to residual oxygen in the annealing vacuum furnace (5\,$\times$10$^{-7}$\,Torr). Our results also suggest the anodized surfaces that produce smoother Nb$_3$Sn rely on increasing the quantity and even distribution of adsorbed SnCl$_2$, and enhancing the reaction kinetics through excessive SnCl$_2$ and Nb oxide sources. These insights into the nucleation mechanisms of intermetallic Nb$_3$Sn alloys will guide the development of optimized growth processes and new synthesis recipes for producing high-quality Nb$_3$Sn superconductors.

% If in two-column mode, this environment will change to single-column format so that long equations can be displayed. 
%\begin{widetext}
%$$\mbox{put long equation here}$$
%\end{widetext}

\begin{acknowledgments}
This work was supported by the U.S. National Science Foundation under Award PHY-1549132, the Center for Bright Beams. This work made use of the Cornell Center for Materials Research Shared Facilities which are supported through the NSF MRSEC program (DMR-1719875). 
\end{acknowledgments}

\section*{Conflicts of Interest}
The authors declare no competing financial interest.

\section*{Data Availability}
Data are available by contacting the corresponding author.

\bibliography{Draft}

\end{document}